# Multiferroic materials based on transition-metal dichalcogenides: Potential platform for reversible control of Dzyaloshinskii-Moriya interaction and skyrmion via electric field

Ziji Shao[1,2], Jinghua Liang[1], Qirui Cui[1], Mairbek Chshiev[3,4], Albert Fert[5], Tiejun Zhou[2,*], Hongxin Yang[1,*]

[1] *Ningbo Institute of Materials Technology and Engineering, Chinese Academy of Sciences, Ningbo 315201, China; Center of Materials Science and Optoelectronics Engineering, University of Chinese Academy of Sciences, Beijing 100049, China*

[2] *College of Electronics and Information, Hangzhou Dianzi University, Hangzhou 310018, China*

[3] *Univ. Grenoble Alpes, CEA, CNRS, Spintec, 38000, Grenoble, France*

[4] *Institut Universitaire de France, 75231 Paris, France*

[5] *Université Paris-Saclay, Unité Mixte de Physique CNRS-Thales, Palaiseau 91767, France*

*Email：tjzhou@hdu.edu.cn

*Email：hongxin.yang@nimte.ac.cn

**Abstract**

Exploring novel two-dimensional multiferroic materials that can realize electric-field control of two-dimensional magnetism has become an emerging topic in spintronics. Using first-principles calculations, we demonstrate that non-metallic bilayer transition metal dichalcogenides (TMDs) can be an ideal platform for building multiferroics by intercalated magnetic atoms. Moreover, we unveil that with Co intercalated bilayer $MoS_2$, $Co(MoS_2)_2$, two energetic degenerate states with opposite chirality of Dzyaloshinskii-Moriya interaction (DMI) are the ground states, indicating electric-field control of the chirality of topologic magnetism such as skyrmions can be realized in this type of materials by reversing the electric polarization. These findings pave the way for electric-field control of topological magnetism in two-dimensional multiferroics with intrinsic magnetoelectric coupling.

The manipulation of spin is the base of spintronic devices with new capabilities and potentially superior performance. [1] Electric control of magnetism is considered to be an efficient approach for applications with low energy consumption and high-speed magnetic switching. Chiral magnetic structures such as chiral domain walls, [2,3] helical structures, [4,5] and magnetic skyrmions [6-8] which generally originate from Dzyaloshinskii–Moriya interactions (DMIs) are highly promising for next generation spintronic devices and are required to be electrically controllable for integration into modern electronic technology. [9-11] It has been reported that electric control of skyrmions can been realized by spin-polarized current based on spin-transfer torque or spin-orbit torque effects. [12-16] However, for industrial applications, ultralow electric current threshold to move domain walls and skyrmions is demanded to reduce energy consumption. [17] It was suggested that the coupling between magnetism and electric polarization (ME) allows the manipulation of magnetization through electric field with low power consumption. [18-23] Two-dimensional (2D) ferromagnetic materials with broken inverse symmetry have provided a promising opportunity to control their DMI. [24-27] Due to ME coupling, the DMI and topological magnetic structure switching mediated by electric field is found in multiferroic $VOI_2$ monolayer [28] and $Ca_3FeOsO_7$ bilayer, [29] but no skyrmions are obtained in these systems. Recently, the transformation between four states of skyrmions controlled by out-of-plane electric field was demonstrated in CrN monolayer theoretically, [24] guiding towards realization of the electric-field manipulation of the topological chirality and promoting further explorations of novel multiferroic materials allowing electric control of skyrmions.

A class of covalently bonded self-intercalated bilayer transition metal dichalcogenides (TMDs) are successfully synthesized recently using both molecular beam epitaxy (MBE) and chemical vapor deposition (CVD) methods. [30] The term ic-2D crystals has been coined for structures such as $TaS_2$-Ta-$TaS_2$ grown by MBE. [30] By varying the coverage of the intercalated metal from low concentration to complete intercalation with the intercalated atoms occupying the octahedral vacancies between the two S layers, the magnetic properties can be tuned. In addition, experimental results have shown that $CoMoS_2$ present ferromagnetic properties with out-of-plane magnetization and Curie temperature $T_c$ around 100 K. [31]

Herein, we propose that the bilayer TMDs can serve as an efficient platform for forming 2D multiferroic systems with intrinsic ME coupling, such as Co intercalated bilayer $MoS_2$. The intercalated Co atoms not only introduce the magnetism but also induce ferroelectricity via formation of polar structure. Therefore, external electric field can alter the magnetic properties when reversing the electric polarization. The first principles calculations suggest two equivalent multiferroic phases of $Co(MoS_2)_2$ (denoted as FE1 and FE2) with opposite chirality of DMI. Moreover, the chirality of skyrmions can be controlled by vertically applied electric field in $Co(MoS_2)_2$.

The first principles calculations are performed in the framework of density-functional theory (DFT) [32] as implemented in the Vienna *ab initio* simulation package (VASP). [33] The electron-ion interaction is described through the projected augmented wave (PAW) method [34-36] and the electronic exchange-correlation functional is treated by the generalized gradient approximation (GGA) parametrized by Perdew,

Burke, and Ernzerhof (PBE). [37] In order to well describe the 3$d$ electrons, we employ the GGA+U method [38] with an effective U$_{eff}$ = 3 eV for 3$d$ orbitals of Co atoms. [39,40] The plane-wave cutoff energy is set to 500 eV. The first Brillouin zone is sampled by a Γ-centered 25 × 25 × 1 k-point mesh. Structural relaxations are performed until the forces become smaller than 0.001 eV/Å per atom. The phonon dispersion is calculated using density functional perturbation theory (DFPT), [41] as implemented in the PHONOPY code. [42,43] The magnetic anisotropy energy (MAE) $K$ can be obtained by calculating the energy difference between in-plane [100] and out-of-plane [001] magnetization orientations. Heisenberg exchange coupling $J$ is calculated by comparing the energies between ferromagnetic and antiferromagnetic states. [25] We employ the chirality-dependent total energy difference approach to calculate the DMI $d$. This method has been successfully applied on Co/Pt film, [44,45] Co/MgO, [45,46] Co/graphene [47] and 2D magnetic materials such as MnXY, CrXY and CrN as well. [24-26] The details are shown in Supporting Information. [48] A Γ-centered 6 × 24 × 1 k-point mesh is adopted in the calculations of DMI. A positive $d$ indicates the anti-clockwise (ACW) spin configurations while a negative $d$ is clockwise (CW) one.

We first systematically investigate all the possible atomic configurations [Fig. S1(a) of the Supplemental Material [48]] of Co(MoS$_2$)$_2$ using DFT calculations. The geometry of Co(MoS$_2$)$_2$ structures can be seen as two MoS$_2$ layers intercalated by one layer of Co atoms. From the calculated total energy dependence as a function of lattice constant for each structure [Fig. S1(b) of the Supplemental Material [48]], we find that the energetic degenerate ferroelectric phases FE1 and FE2 are the ground states. The dynamical stability is confirmed through phonon calculation and *ab initio* molecular dynamic simulation [Fig. S2 of the Supplemental Material [48]]. The structures of FE1 and FE2 phases are depicted in Fig. 1(a) and Fig. 1(b) with optimized lattice constant 3.224 Å. Each Co atom is tetra-coordinate and occupies the trigonal-pyramidal vacancies between two stacking $MoS_2$ layers. The electron localization functions (ELFs) [Fig. S3 of the Supplemental Material [48]] indicate the mixture of covalent and ionic interactions between Co and the adjacent S. This tetra-coordinate environment causes different vertical distances between Co atoms and two adjacent $MoS_2$ monolayers, which induces the electric polarization. The electric dipoles of the stable ferroelectric state FE1 and FE2 are calculated to be 0.029 e Å/f.u. and -0.029 e Å/f.u., respectively. The reversion of the electric polarization can be realized via the displacement of Co atom. [53] The ferroelectric switching pathways are investigated by using the nudged elastic band (NEB) method. [54] Here, we adopt a tow-step switching path with a meta-stable phase being the transition state. In many theoretical studies, this kind of multistep switching path can efficiently reduce the energy of barriers. [55] As shown in Fig. 1(c), the FE1 phase first convert into the transition phase via shifting the top four atomic layers along diagonal as illustrated by red arrows. Then the top $MoS_2$ monolayer of the transition phase moves back to the original position to realize the transformation from the transition phase to FE2 phase. The overall switching barrier is calculated to be 0.485 eV/f.u. [Fig. 1(c)].

The partial electronic density of states [PDOS in Fig. 2] provides two important information of $Co(MoS_2)_2$ monolayer. First, the existence of electronic density at the Fermi level indicates a metallic characteristic for ferroelectric $Co(MoS_2)_2$. Therefore, the electrons are confined within the slab and only vertical polarization exists. [56] Second, $Co(MoS_2)_2$ is ferromagnetic, which is mainly induced by intercalated Co atoms. The magnetic moment is calculated to be about 1.28 $\mu_B$ per Co atom. The atomic-projected electronic density of states [Fig. 2(b)] and spin density [insert in Fig. 2(b)] show that the magnetism is mainly caused by the partial occupation of $d_{z^2}$, $d_{xy}$ and $d_{x^2-y^2}$ states of Co atoms. Since the spin polarization is mainly provided by Co atoms, the following Hamiltonian model is employed in systems considered here for further investigating the magnetic properties:

$$H = -J\sum_{i,j} \mathbf{S}_j \cdot \mathbf{S}_j - K\sum_i (S_i^z)^2 - \sum_{i,j} \mathbf{D}_{i,j} \cdot (\mathbf{S}_i \times \mathbf{S}_j) - \mu_{Co} B \sum_i S_i^z, \qquad (1)$$

where $\mathbf{S}_i$ indicates spin unit vector of the *ith* Co atom. Here, we only take the magnetic coupling between the nearest Co atoms into consideration, therefore *i,j* represent the nearest neighbor Co atom pairs. The first three terms correspond to DMI, Heisenberg exchange coupling, and single-ion anisotropy, characterized by $\mathbf{D}_{i,j}$, *J*, and *K*, respectively. The specific calculation method is given in Supporting Information. [48] The last term represents the Zeeman interaction with $\mu_{Co}$ and *B* being magnetic moment of the Co atoms and external magnetic field, respectively. The magnetic ground states of ferroelectric $Co(MoS_2)_2$ is calculated to be ferromagnetic out-of-plane with the nearest neighbor magnetic interaction parameter *J* about 5.65 meV and the PMA about 0.09 meV, respectively. The broken of inversion symmetry induced by tetra-coordinate Co atoms provides an opportunity for the appearance of chiral domain wall or skyrmions in case of significant spin-orbit coupling (SOC) energy sources present. [57,58]

Therefore, we investigated the DMI and the associated SOC energy difference $\Delta E_{SOC}$ of these two equivalent ferroelectric phases. The DMI strengths for FE1 and FE2 phases with up and down electric polarization are the same with values of 1.34 meV. However, their chirality is opposite being ACW for FE1 and CW for FE2, respectively. As shown in Fig. 3(a) and Fig. 3(c), the largest associated $\Delta E_{SOC}$ originates from Co layer and located away Mo layer with both contributing equally. These results indicate that both Fert-Levy [44,59] and Rashba [47] physical mechanisms governing the DMI coexist in the Co intercalated multiferroic phase proposed here. In order to explore the influence of Rashba SOC on the DMI, [60-63] the electronic band structures with magnetization pointing along [210] and [$\bar{2}\bar{1}0$] directions are calculated with the SOC switched on. As shown in Fig. 3(b) and Fig. 3(d), the significant Rashba-type **k** dependent splitting indicates large Rashba effect in this system. To quantitatively estimate its contribution to the total DMI, the Rahba-type DMI parameter $D_R$ is estimated with $D_R = \left(\frac{4m^*A}{\hbar^2}\right)\alpha_R$, [60] where $m^*$ indicates the effective mass of electron, $\alpha_R$ is the Rashba coefficient which can be extracted from equation $\alpha_R = 2\Delta E/\Delta k$, and A represents the exchange stiffness estimated to be 2.45 meV Å$^2$/f.u. in Co(MoS$_2$)$_2$. [64] Here we mainly focus on three bands [1, 2 and 3 shown in Fig. 3(b) and Fig. 3(d)] around M point along K-M-K' path under the Fermi level. In terms of the oblivious bands shifts, the Rashba coefficient $\alpha_R$ is estimated to be 0.779, 1.048 and 0.908 eV Å for band 1, 2 and 3 with effective masses $m^*$ equal to 5.07, -8.27 and 5.65 in units of the rest mass of electron $m_e$, respectively. The corresponding magnitudes of $D_R$ evaluated from these parameters are 5.17, -11.35 and 6.74 meV Å/f.u. This allows estimating the total $D_R$ resulting from these three bands being about 0.56 meV Å/f.u. The calculated Rashba-type $D_R$ is comparable to the half of the value obtained from DFT calculations, allowing to conclude that the Rashba SOC has the comparable contribution to the DMI with the Fert-Levy SOC. The existence of the sizable Rashba-type DMI in the ferroelectric $Co(MoS_2)_2$ phase indicates the possibility to realize electric-field control of the chirality of topological spin structures. As shown in Fig. 3, the transition from FE1 to FE2 phase accompanied with the inversion of electric polarization induces the inversion of the chirality of DMI from ACW to CW due to the inversion of $\Delta E_{SOC}$ and the Rashba splitting. The results show that the electric control on the chirality of chiral magnetic structures can be realized in our multiferroic $Co(MoS_2)_2$ materials. Using all magnetic parameters estimated from first principles calculations listed in Table SI of the Supplemental Material [48] we performed micromagnetic simulations with 100 nm×100 nm scale using mumax3 package [65] for two ferroelectric phases FE1 and FE2 with ACW and CW DMI under different external magnetic fields, respectively. Ferromagnetic states with wormlike domains are found under 0 T magnetic field. When the magnetic field is applied, the skyrmions are induced in the range of 1-5 T, and then disappear under 6 T [Fig. 4(a)]. Furthermore, the chirality of wormlike domains and skyrmions can be switched between CW and ACW configurations through application of external electric field since the reversing of electric polarization allows controlling the DMI chirality [Fig. S5 of the Supplemental Material [48]]. With the polarity of skyrmions **M** defined as magnetization directions in skyrmion core, we thus obtain four topological configurations interchangeable and controlled through application of external electric and magnetic fields in $Co(MoS_2)_2$ [Fig. 4(b)].

In addition to $Co(MoS_2)_2$, we also investigated other potential candidates for the proposed ME coupling including $Co(MoSe_2)_2$, $Co(MoTe_2)_2$ and $Co(WS_2)_2$ as summarized in Table SII of the Supplemental Material [48]. All of them are ferroelectric with electronic polarizations about 0.012, 0.012 and 0.023 e Å/f.u. Both $Co(MoTe_2)_2$ and $Co(WS_2)_2$ have considerable DMI amplitudes of about 1.044 and 2.632 meV. These values are comparable to representative FM/HM heterostructures, such as Fe/Ir (111) (~1.7 meV) [66] and Co/Pt (~3.0 meV) [44] thin films, and the 2D magnetoelectric multiferroics CrN (~3.74 meV) [26] in which the DMI generated either the Fert-Levy or Rashba effect mechanism, respectively. Skyrmion states are found to exist in these systems. Even for the $Co(MoSe_2)_2$ with the weakest DMI interaction (~0.440 meV), the value is comparable to the Rashba-type DMI in the graphene/Co system (~0.16 meV) [47] with chiral domain walls. Therefore, Co intercalated TMD systems studied here are expected to be potential multiferroic materials with realizable chirality switching via electric field.

In conclusion, we proposed bilayer TMDs as a platform towards realization of novel class of 2D multiferroic materials via intercalating magnetic atoms. Intrinsic multiferroicity with ME coupling can be obtained with intercalated magnetic atoms located at the positions with different interlayer spacing between two TMD layers. Using first principles calculations, the Co intercalated bilayer $MoS_2$, $Co(MoS_2)_2$ is found to be multiferroic with degenerate DMI for two ferroelectric states. As a result, the chirality of skyrmions can be reversed via electric field thanks to the ME coupling. We also shown that these properties can be realized in other multiferroic TMDs with Co intercalated within bilayer $MoTe_2$ and $WS_2$ systems, i.e. $Co(MoTe_2)_2$ and $Co(WS_2)_2$. In addition, different combinations of the TMD layers can further enrich the varieties of this kind of multiferroic materials [26] like $Co(Mo_2SSe)$. Considering experimentally successful synthesis of the self-intercalated TMDs with well-developed intercalation techniques [30] and the as-exfoliated nano-sheets $AgCrS_2$ with a general redox-controlled strategy [67] proposed recently, the multiferroic $Co(MoS)_2$ provides a new route towards novel two-dimensional multiferroic materials that can realize the electric control of topological magnetism in 2D spintronics.

# Figure and figure captions

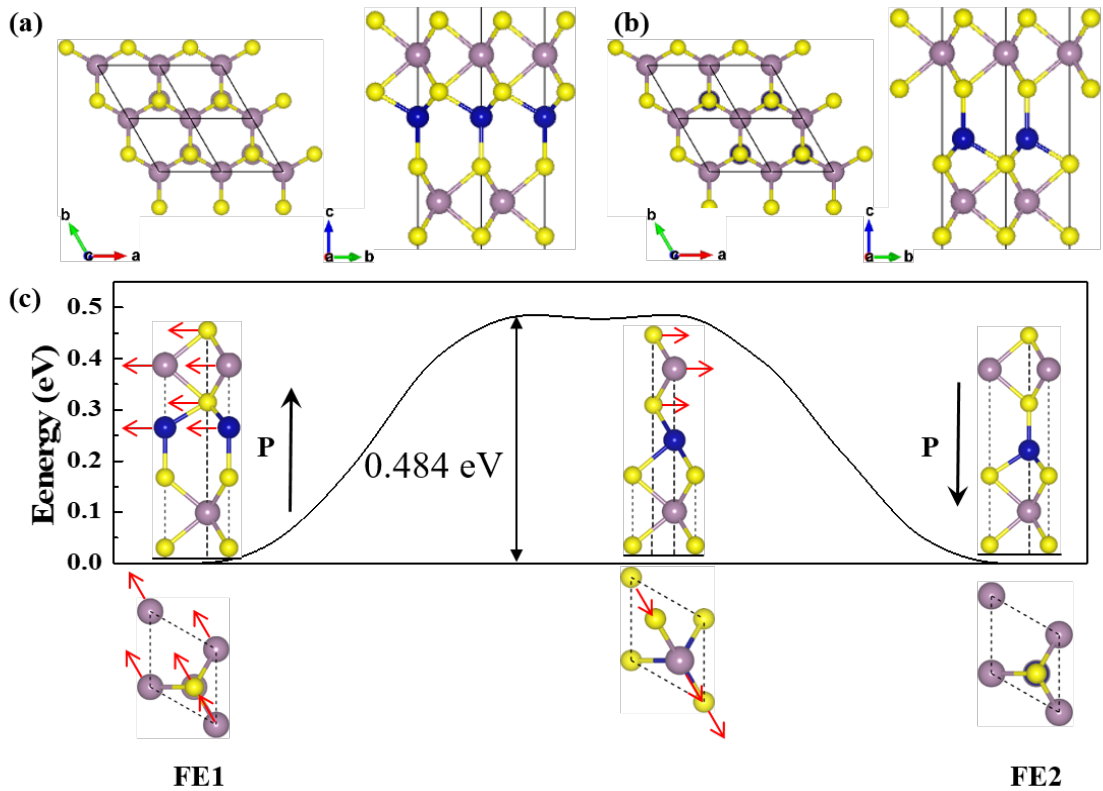

FIG. 1. The top (left) and side (right) views of (a) FE1 and (b) FE2 phases of Co(MoS$_2$)$_2$ monolayer. Blue, pink and yellow balls represent Co, Mo and S atoms, respectively. (c) The ferroelectric switching pathway of Co(MoS$_2$)$_2$ by NEB method with the top and side view of transition structure during the ferroelectric switching. The red arrows depicted the orientation of atomic movement.

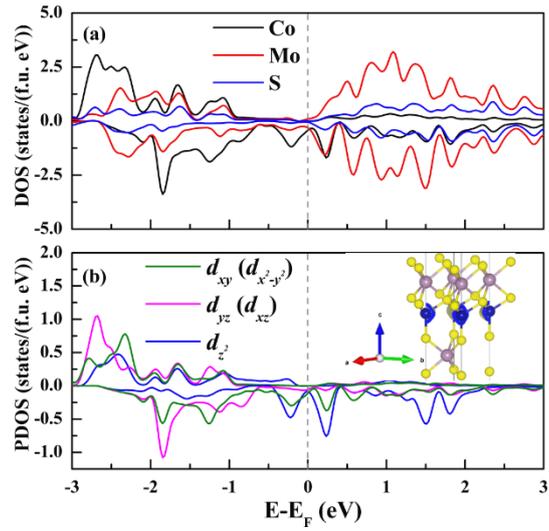

FIG. 2. (a) The atom-projected partial DOS for Co(MoS$_2$)$_2$ monolayer. (b) The atomic orbital-projected DOS for each $d$ orbital of Co atom in Co(MoS$_2$)$_2$ monolayer. The spin density of Co(MoS$_2$)$_2$ monolayer is shown in insert, the blue, pink and yellow balls represent Co, Mo and S atoms, respectively.

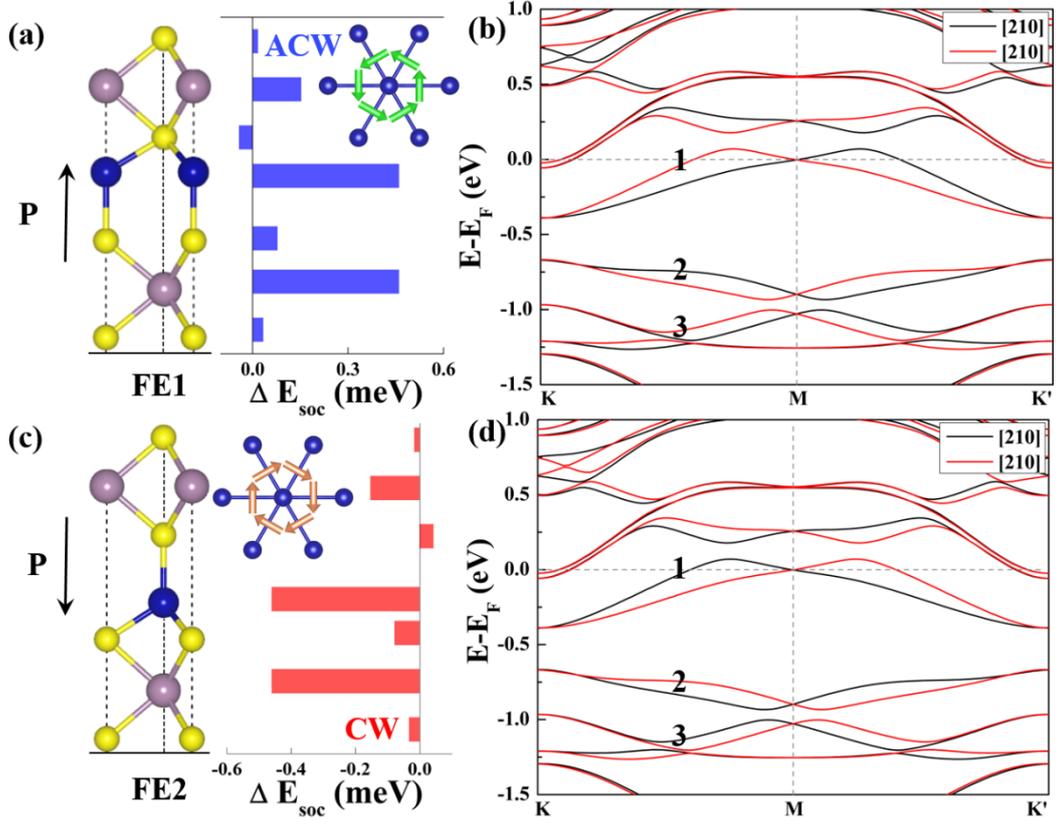

FIG. 3. Atom-resolved localization of the SOC energy $\Delta E_{SOC}$ in Co(MoS$_2$)$_2$ monolayer with (a) up and (c) down electric polarization. Band structures of (b) up and (d) down polarized Co(MoS$_2$)$_2$ monolayer with the magnetization direction along [210] (black lines) and [$\bar{2}\bar{1}0$] (red lines). Three bands under the Fermi level (band 1, 2 and 3) are chosen to estimate $D_R$ value.

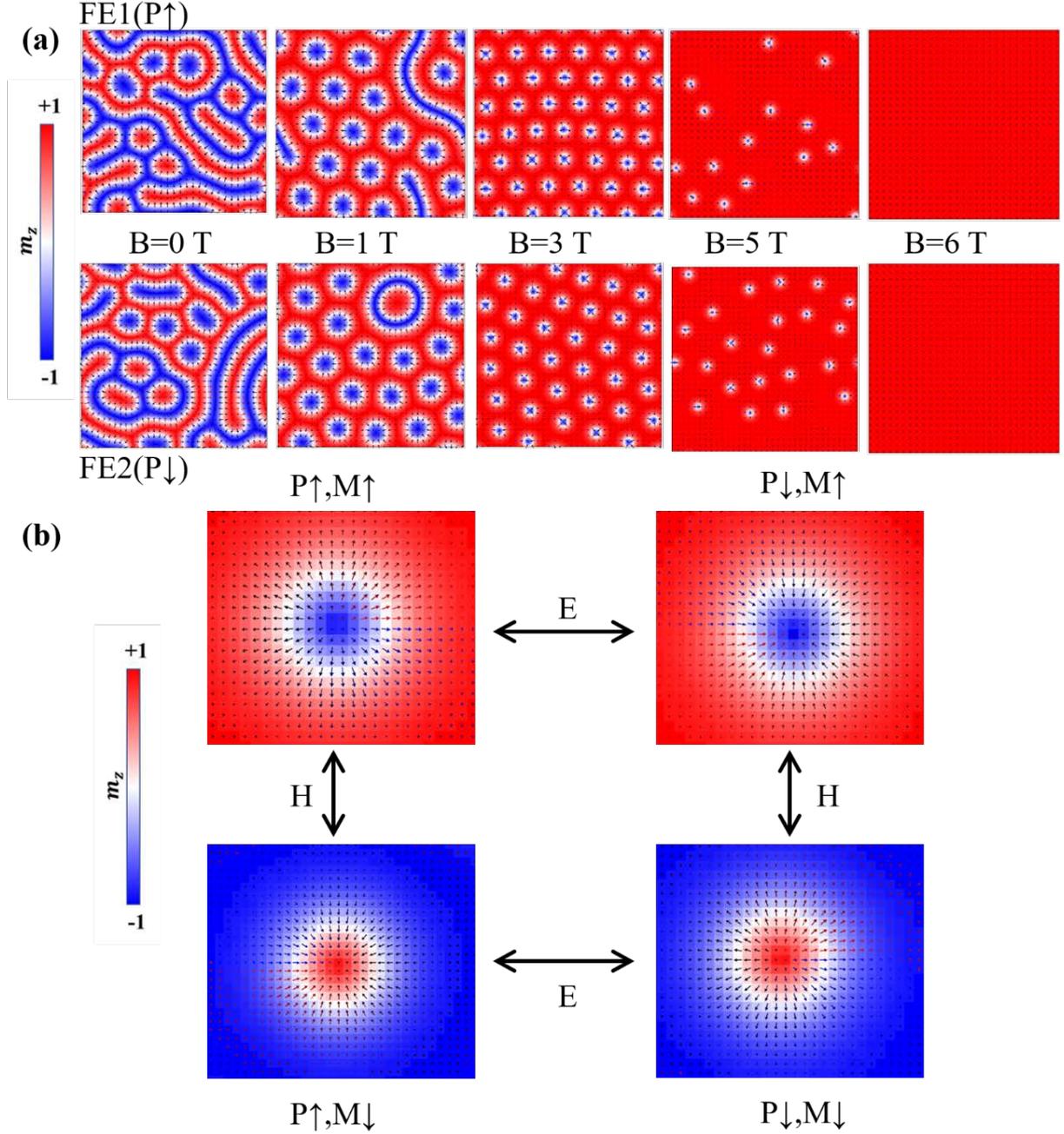

FIG. 4. (a) Spin textures for Co(MoS$_2$)$_2$ in real-space under 0, 1, 3, 5, and 6 T in FE1 and FE2 phases. The color map indicates the out-of-plane spin component of Co atoms. Panel (b) illustrates the realization of the switching of four types of skyrmions (P↑,M↑), (P↑,M↓), (P↓,M↑), and (P↓,M↓) with different chirality and polarity via external field, similar as that in CrN. [24] M and P represent the magnetization and electric polarization, respectively. ACW and CW are anti-clockwise and clockwise.

# Supporting Information

Additional information about structures and stabilities for Co(MoS$_2$)$_2$, calculations methods of the magnetic interaction parameters *J*, *K*, and *D*, the evolution of the magnetic parameters and spin textures under external electric field, the impact of effective Coulomb interaction U$_{eff}$, magnetic parameters of two ferroelectric phases in micromagnetic simulations and other potential TMDs based multiferroic materials. (PDF)

# Acknowledgements

This work was supported by "Pioneer" and "Leading Goose" R&D Program of Zhejiang Province under Grant No. 2022C01053; National Natural Science Foundation of China (Grant Nos. 11874059 and 12174405); Ningbo Key Scientific and Technological Project (Grant No. 2021000215); Zhejiang Provincial Natural Science Foundation (Grant No. LR19A040002); Beijing National Laboratory for Condensed Matter Physics (No.2021000123); and European Union's Horizon 2020 research and innovation Program under grant agreement 881603 (Graphene Flagship).

# References

[1] S. S. P. Parkin, M. Hayashi, and L. Thomas, Magnetic Domain-Wall Racetrack Memory, Science **320**, 190 (2008).

[2] A. Thiaville, S. Rohart, É. Jué, V. Cros, and A. Fert, Dynamics of Dzyaloshinskii domain walls in ultrathin magnetic films, Europhys. Lett. **100**, 57002 (2012).

[3] K.-S. Ryu, L. Thomas, S.-H. Yang, and S. Parkin, Chiral spin torque at magnetic domain walls, Nat. Nanotechnol. **8**, 527 (2013).

[4] M. Bode, M. Heide, K. von Bergmann, P. Ferriani, S. Heinze, G. Bihlmayer, A. Kubetzka, O. Pietzsch, S. Blügel, and R. Wiesendanger, Chiral magnetic order at surfaces driven by inversion asymmetry, Nature **447**, 190 (2007).

[5] C. S. Spencer, J. Gayles, N. A. Porter, S. Sugimoto, Z. Aslam, C. J. Kinane, T. R. Charlton, F. Freimuth, S.

Chadov, S. Langridge, J. Sinova, C. Felser, S. Blügel, Y. Mokrousov, and C. H. Marrows, Helical magnetic structure and the anomalous and topological Hall effects in epitaxial B20 Fe$_{1-y}$Co$_y$Ge films, Phys. Rev. B **97**, 214406 (2018).

[6] A. Bogdanov and D. Yablonskiui, Thermodynamically stable "vortices" in magnetically ordered crystals. The mixed state of magnets, Sov. Phys. JETP **68**, 101 (1989).

[7] S. Mühlbauer, B. Binz, F. Jonietz, C. Pfleiderer, A. Rosch, A. Neubauer, R. Georgii, and P. Böni, Skyrmion Lattice in a Chiral Magnet, Science **323**, 915 (2009).

[8] X. Z. Yu, N. Kanazawa, Y. Onose, K. Kimoto, W. Z. Zhang, S. Ishiwata, Y. Matsui, and Y. Tokura, Near room-temperature formation of a skyrmion crystal in thin-films of the helimagnet FeGe, Nat. Mater. **10**, 106 (2011).

[9] A. Fert, N. Reyren, and V. Cros, Magnetic skyrmions: advances in physics and potential applications, Nat. Rev. Mater. **2**, 17031 (2017).

[10] F. Matsukura, Y. Tokura, and H. Ohno, Control of magnetism by electric fields, Nat. Nanotechnol. **10**, 209 (2015).

[11] A. C. Garcia-Castro, W. Ibarra-Hernandez, E. Bousquet, and A. H. Romero, Direct Magnetization-Polarization Coupling in BaCuF$_4$, Phys. Rev. Lett. **121**, 117601 (2018).

[12] A. Fert, V. Cros, and J. Sampaio, Skyrmions on the track, Nat. Nanotechnol. **8**, 152 (2013).

[13] A. Rosch, Moving with the current, Nat. Nanotechnol. **8**, 160 (2013).

[14] Y. Zhou, Magnetic skyrmions: intriguing physics and new spintronic device concepts, Nat. Sci. Rev. **6**, 210 (2018).

[15] W. Jiang, P. Upadhyaya, W. Zhang, G. Yu, M. B. Jungfleisch, F. Y. Fradin, J. E. Pearson, Y. Tserkovnyak, K. L. Wang, O. Heinonen, S. G. E. te Velthuis, and A. Hoffmann, Blowing magnetic skyrmion bubbles, Science **349**, 283 (2015).

[16] S. Woo, K. Litzius, B. Krüger, M.-Y. Im, L. Caretta, K. Richter, M. Mann, A. Krone, R. M. Reeve, M. Weigand, P. Agrawal, I. Lemesh, M.-A. Mawass, P. Fischer, M. Kläui, and G. S. D. Beach, Observation of room-temperature magnetic skyrmions and their current-driven dynamics in ultrathin metallic ferromagnets, Nat. Mater. **15**, 501 (2016).

[17] F. Jonietz, S. Mühlbauer, C. Pfleiderer, A. Neubauer, W. Münzer, A. Bauer, T. Adams, R. Georgii, P. Böni, R. A. Duine, K. Everschor, M. Garst, and A. Rosch, Spin Transfer Torques in MnSi at Ultralow Current Densities, Science **330**, 1648 (2010).

[18] P. Huang, M. Cantoni, A. Kruchkov, J. Rajeswari, A. Magrez, F. Carbone, and H. M. Rønnow, In Situ Electric

Field Skyrmion Creation in Magnetoelectric $Cu_2OSeO_3$, Nano Lett. **18**, 5167 (2018).

[19] C. Ma, X. Zhang, J. Xia, M. Ezawa, W. Jiang, T. Ono, S. N. Piramanayagam, A. Morisako, Y. Zhou, and X. Liu, Electric Field-Induced Creation and Directional Motion of Domain Walls and Skyrmion Bubbles, Nano Lett. **19**, 353 (2019).

[20] J. S. White, K. Prša, P. Huang, A. A. Omrani, I. Živković, M. Bartkowiak, H. Berger, A. Magrez, J. L. Gavilano, G. Nagy, J. Zang, and H. M. Rønnow, Electric-Field-Induced Skyrmion Distortion and Giant Lattice Rotation in the Magnetoelectric Insulator $Cu_2OSeO_3$, Phys. Rev. Lett. **113**, 107203 (2014).

[21] P.-J. Hsu, A. Kubetzka, A. Finco, N. Romming, K. von Bergmann, and R. Wiesendanger, Electric-field-driven switching of individual magnetic skyrmions, Nat. Nanotechnol. **12**, 123 (2017).

[22] T. Srivastava, M. Schott, R. Juge, V. Křižáková, M. Belmeguenai, Y. Roussigné, A. Bernand-Mantel, L. Ranno, S. Pizzini, S.-M. Chérif, A. Stashkevich, S. Auffret, O. Boulle, G. Gaudin, M. Chshiev, C. Baraduc, and H. Béa, Large-Voltage Tuning of Dzyaloshinskii–Moriya Interactions: A Route toward Dynamic Control of Skyrmion Chirality, Nano Lett. **18**, 4871 (2018).

[23] Y. Wang, L. Wang, J. Xia, Z. Lai, G. Tian, X. Zhang, Z. Hou, X. Gao, W. Mi, C. Feng, M. Zeng, G. Zhou, G. Yu, G. Wu, Y. Zhou, W. Wang, X.-x. Zhang, and J. Liu, Electric-field-driven non-volatile multi-state switching of individual skyrmions in a multiferroic heterostructure, Nat. Commun. **11**, 3577 (2020).

[24] J. Liang, Q. Cui, and H. Yang, Electrically switchable Rashba-type Dzyaloshinskii-Moriya interaction and skyrmion in two-dimensional magnetoelectric multiferroics, Phys. Rev. B **102**, 220409 (2020).

[25] Q. Cui, J. Liang, Z. Shao, P. Cui, and H. Yang, Strain-tunable ferromagnetism and chiral spin textures in two-dimensional Janus chromium dichalcogenides, Phys. Rev. B **102**, 094425 (2020).

[26] J. Liang, W. Wang, H. Du, A. Hallal, K. Garcia, M. Chshiev, A. Fert, and H. Yang, Very large Dzyaloshinskii-Moriya interaction in two-dimensional Janus manganese dichalcogenides and its application to realize skyrmion states, Phys. Rev. B **101**, 184401 (2020).

[27] Q. Cui, J. Liang, B. Yang, Z. Wang, P. Li, P. Cui, and H. Yang, Giant enhancement of perpendicular magnetic anisotropy and induced quantum anomalous Hall effect in graphene/$NiI_2$ heterostructures via tuning the van der Waals interlayer distance, Phys. Rev. B **101**, 214439 (2020).

[28] C. Xu, P. Chen, H. Tan, Y. Yang, H. Xiang, and L. Bellaiche, Electric-Field Switching of Magnetic Topological Charge in Type-I Multiferroics, Phys. Rev. Lett. **125**, 037203 (2020).

[29] J. Zhang, X. Shen, Y. Wang, C. Ji, Y. Zhou, J. Wang, F. Huang, and X. Lu, Design of Two-Dimensional

Multiferroics with Direct Polarization-Magnetization Coupling, Phys. Rev. Lett. **125**, 017601 (2020).

[30] X. Zhao, P. Song, C. Wang, A. C. Riis-Jensen, W. Fu, Y. Deng, D. Wan, L. Kang, S. Ning, J. Dan, T. Venkatesan, Z. Liu, W. Zhou, K. S. Thygesen, X. Luo, S. J. Pennycook, and K. P. Loh, Engineering covalently bonded 2D layered materials by self-intercalation, Nature **581**, 171 (2020).

[31] M. Huang, J. Xiang, C. Feng, H. Huang, P. Liu, Y. Wu, A. T. N'Diaye, G. Chen, J. Liang, H. Yang, J. Liang, X. Cui, J. Zhang, Y. Lu, K. Liu, D. Hou, L. Liu, and B. Xiang, Direct Evidence of Spin Transfer Torque on Two-Dimensional Cobalt-Doped $MoS_2$ Ferromagnetic Material, ACS Appl. Electron. Mater. **2**, 1497 (2020).

[32] W. Kohn and L. J. Sham, Self-Consistent Equations Including Exchange and Correlation Effects, Phys. Rev. **140**, A1133 (1965).

[33] G. Kresse and J. Furthmüller, Efficient iterative schemes for ab initio total-energy calculations using a plane-wave basis set, Phys. Rev. B **54**, 11169 (1996).

[34] G. Kresse and J. Hafner, Ab initio molecular dynamics for liquid metals, Phys. Rev. B **47**, 558 (1993).

[35] G. Kresse and J. Hafner, Ab initio molecular-dynamics simulation of the liquid-metal-amorphous-semiconductor transition in germanium, Phys. Rev. B **49**, 14251 (1994).

[36] G. Kresse and J. Furthmüller, Efficiency of ab-initio total energy calculations for metals and semiconductors using a plane-wave basis set, Comput. Mater. Sci. **6**, 15 (1996).

[37] J. P. Perdew, K. Burke, and M. Ernzerhof, Generalized Gradient Approximation Made Simple, Phys. Rev. Lett. **77**, 3865 (1996).

[38] V. I. Anisimov, F. Aryasetiawan, and A. I. Lichtenstein, First-principles calculations of the electronic structure and spectra of strongly correlated systems: the LDA+U method, J. Phys. Condens. Matter **9**, 767 (1997).

[39] L. Wang, T. Maxisch, and G. Ceder, Oxidation energies of transition metal oxides within the GGA+U framework, Phys. Rev. B **73**, 195107 (2006).

[40] C. Le, S. Qin, and J. Hu, Electronic physics and possible superconductivity in layered orthorhombic cobalt oxychalcogenides, Sci. Bull. **62**, 563 (2017).

[41] S. Baroni, S. de Gironcoli, A. Dal Corso, and P. Giannozzi, Phonons and related crystal properties from density-functional perturbation theory, Rev. Mod. Phys. **73**, 515 (2001).

[42] A. Togo, F. Oba, and I. Tanaka, First-principles calculations of the ferroelastic transition between rutile-type and $CaCl_2$-type $SiO_2$ at high pressures, Phys. Rev. B **78**, 134106 (2008).

[43] A. Togo and I. Tanaka, First principles phonon calculations in materials science, Scr. Mater. **108**, 1 (2015).

[44] H. Yang, A. Thiaville, S. Rohart, A. Fert, and M. Chshiev, Anatomy of Dzyaloshinskii-Moriya Interaction at Co/Pt Interfaces, Phys. Rev. Lett. **115**, 267210 (2015).

[45] H. Yang, O. Boulle, V. Cros, A. Fert, and M. Chshiev, Controlling Dzyaloshinskii-Moriya Interaction via Chirality Dependent Atomic-Layer Stacking, Insulator Capping and Electric Field, Sci. Rep. **8**, 12356 (2018).

[46] O. Boulle, J. Vogel, H. Yang, S. Pizzini, D. de Souza Chaves, A. Locatelli, T. O. Menteş, A. Sala, L. D. Buda-Prejbeanu, O. Klein, M. Belmeguenai, Y. Roussigné, A. Stashkevich, S. M. Chérif, L. Aballe, M. Foerster, M. Chshiev, S. Auffret, I. M. Miron, and G. Gaudin, Room-temperature chiral magnetic skyrmions in ultrathin magnetic nanostructures, Nat. Nanotechnol. **11**, 449 (2016).

[47] H. Yang, G. Chen, A. A. C. Cotta, A. T. N'Diaye, S. A. Nikolaev, E. A. Soares, W. A. A. Macedo, K. Liu, A. K. Schmid, A. Fert, and M. Chshiev, Significant Dzyaloshinskii–Moriya interaction at graphene–ferromagnet interfaces due to the Rashba effect, Nat. Mater. **17**, 605 (2018).

[48] See Supplemental Material at [URL will be inserted by the production group], which includes Refs. [31,39,40,49-52], for computational details of the structures and their stabilities, ELF, calculation methods of magnetic interaction parameters $J$, $K$, and $D$, the evolution of the spin textures under an electric field, the influence of U$_{eff}$ on the magnetic parameters, magnetic parameters for micromagnetic simulations, the lattice constant, magnetic parameters and electric polarization of other intercalted TMD bilayer multiferroelectric materials $Co(MoSe_2)_2$, $Co(MoTe_2)_2$ and $Co(WS_2)_2$.

[49] Y. Wang, J. Lv, L. Zhu, and Y. Ma, CALYPSO: A method for crystal structure prediction, Comput. Phys. Commun. **183**, 2063 (2012).

[50] D. Zhou, D. V. Semenok, D. Duan, H. Xie, W. Chen, X. Huang, X. Li, B. Liu, A. R. Oganov, and T. Cui, Superconducting praseodymium superhydrides, Sci. Adv. **6**, eaax6849 (2020)

[51] M. Yu, X. Liu, and W. Guo, Novel two-dimensional ferromagnetic semiconductors: Ga-based transition-metal trichalcogenide monolayers, Phys. Chem. Chem. Phys. **20**, 6374 (2018).

[52] P. Jiang, L. Kang, X. Zheng, Z. Zeng, and S. Sanvito, Computational prediction of a two-dimensional semiconductor $SnO_2$ with negative Poisson's ratio and tunable magnetism by doping, Phys. Rev. B **102**, 195408 (2020).

[53] T. Zhong, X. Li, M. Wu, and J.-M. Liu, Room-temperature multiferroicity and diversified magnetoelectric couplings in 2D materials, Nat. Sci. Rev. **7**, 373 (2020).

[54] G. Henkelman, B. P. Uberuaga, and H. Jónsson, A climbing image nudged elastic band method for finding saddle

points and minimum energy paths, J. Chem. Phys. **113**, 9901 (2000).

[55] W. Ding, J. Zhu, Z. Wang, Y. Gao, D. Xiao, Y. Gu, Z. Zhang, and W. Zhu, Prediction of intrinsic two-dimensional ferroelectrics in $In_2Se_3$ and other III2-VI3 van der Waals materials, Nat. Commun. **8**, 14956 (2017).

[56] W. Luo, K. Xu, and H. Xiang, Two-dimensional hyperferroelectric metals: A different route to ferromagnetic-ferroelectric multiferroics, Phys. Rev. B **96**, 235415 (2017).

[57] I. Dzyaloshinsky, A thermodynamic theory of "weak" ferromagnetism of antiferromagnetics, J. Phys. Chem. Solids **4**, 241 (1958).

[58] T. Moriya, Anisotropic Superexchange Interaction and Weak Ferromagnetism, Phys. Rev. **120**, 91 (1960).

[59] A. Fert and P. M. Levy, Role of Anisotropic Exchange Interactions in Determining the Properties of Spin-Glasses, Phys. Rev. Lett. **44**, 1538 (1980).

[60] K.-W. Kim, H.-W. Lee, K.-J. Lee, and M. D. Stiles, Chirality from Interfacial Spin-Orbit Coupling Effects in Magnetic Bilayers, Phys. Rev. Lett. **111**, 216601 (2013).

[61] H. Imamura, P. Bruno, and Y. Utsumi, Twisted exchange interaction between localized spins embedded in a one- or two-dimensional electron gas with Rashba spin-orbit coupling, Phys. Rev. B **69**, 121303 (2004).

[62] A. Kundu and S. Zhang, Dzyaloshinskii-Moriya interaction mediated by spin-polarized band with Rashba spin-orbit coupling, Phys. Rev. B **92**, 094434 (2015).

[63] I. A. Ado, A. Qaiumzadeh, R. A. Duine, A. Brataas, and M. Titov, Asymmetric and Symmetric Exchange in a Generalized 2D Rashba Ferromagnet, Phys. Rev. Lett. **121**, 086802 (2018).

[64] H. Kronmuller, H. Kronmüller, and M. Fähnle, Micromagnetism and the microstructure of ferromagnetic solids (Cambridge university press, 2003).

[65] A. Vansteenkiste and B. Van de Wiele, MuMax: A new high-performance micromagnetic simulation tool, J. Magn. Magn. Mater. **323**, 2585 (2011).

[66] B. Dupé, M. Hoffmann, C. Paillard, and S. Heinze, Tailoring magnetic skyrmions in ultra-thin transition metal films, Nat. Commun. **5**, 4030 (2014)

[67] J. Peng, Y. Liu, H. Lv, Y. Li, Y. Lin, Y. Su, J. Wu, H. Liu, Y. Guo, and Z. Zhuo, Stoichiometric two-dimensional non-van der Waals $AgCrS_2$ with superionic behaviour at room temperature, Nat. Chem. **13**, 1235 (2021).

Supporting Information for

Multiferroic materials based on transition-metal dichalcogenides: Potential platform for reversible control of Dzyaloshinskii-Moriya interaction and skyrmion via electric field

Ziji Shao[1,2], Jinghua Liang[1], Qirui Cui[1], Mairbek Chshiev[3,4], Albert Fert[5], Tiejun Zhou[2,*], Hongxin Yang[1,*]

[1]*Ningbo Institute of Materials Technology and Engineering, Chinese Academy of Sciences, Ningbo 315201, China; Center of Materials Science and Optoelectronics Engineering, University of Chinese Academy of Sciences, Beijing 100049, China*

[2]*College of Electronics and Information, Hangzhou Dianzi University, Hangzhou 310018, China*

[3] *Univ. Grenoble Alpes, CEA, CNRS, Spintec, 38000, Grenoble, France*

[4]*Institut Universitaire de France, 75231 Paris, France*

[5] *Université Paris-Saclay, Unité Mixte de Physique CNRS-Thales, Palaiseau 91767, France*

*Email：tjzhou@hdu.edu.cn

*Email：hongxin.yang@nimte.ac.cn

# 1 Structure and stabilities

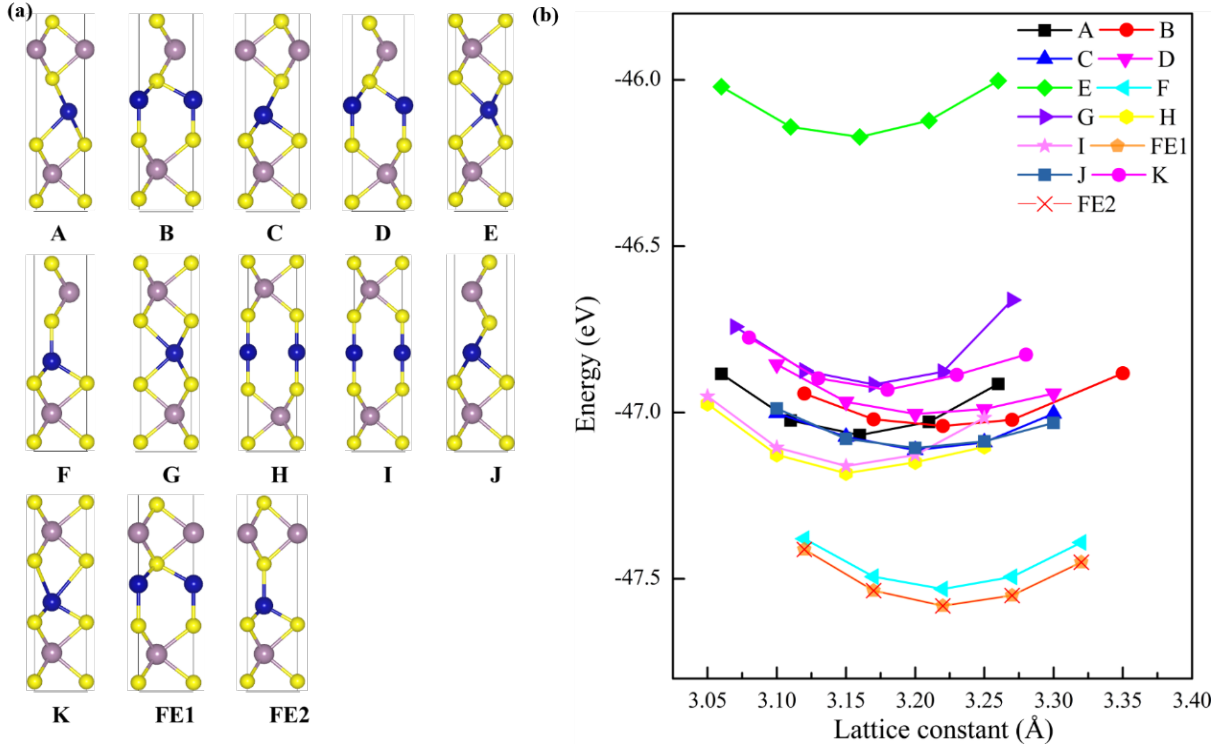

FIG. S1. (a) Side views of all structures we have taken into consideration in Co intercalated bilayer MoS$_2$ system. (b) Calculated total energy as a function of lattice constant for all the considered structures listed in (a).

The calculation results are also examined by the structure search with the particle swarm optimization (PSO) method, as implemented in the CALYPSO code[1]. In our simulation, the best 60% of structures were selected through particle swarm optimization (PSO) to generate the next generation, 40% of structures were generated randomly to guarantee structural diversity. We set the population size to 50 and the number of generations to 30. The low energy of the multiferromagnetic Co(MoS$_2$)$_2$ is confirmed by the structure search results.

The thermodynamic stability is also investigated. First, we calculated the formation energy of the Co(MoS$_2$)$_2$ with respect to the elemental Co, Mo, and S defined as $\Delta E_1 = $ $(E_{(MoS_2)_2Co} - 2E_{Mo} - 4E_S - E_{Co})/n$. We find out that $\Delta E_1$ is -0.306 eV/atom, the negative value indicates that Co(MoS$_2$)$_2$ is thermodynamically stable compared with the elemental Mo, S, and Co. Then, we calculated the formation energy of the Co(MoS$_2$)$_2$ with respect to the bilayer MoS$_2$ and elemental Co, defined as $\Delta E_2 = (E_{(MoS_2)_2Co} - E_{(MoS_2)_2} - E_{Co})/n$. The result is about 0.635 eV/atom and the Co(MoS$_2$)$_2$ is metastable compared with the bilayer MoS$_2$ and elemental Co. However, metastability does not the obstacle to the experimental synthesis of the materials. The inorganic crystal structure database shows that 20% of experimentally synthesized materials are metastable[2]. The formation energy $\Delta E_2$ in our calculation is comparable to the two dimensional experimental synthesized CrGeTe$_3$ (~0.57 eV/atom) [3] and SnO (> 1.4 eV/atom) [4]. In addition, the CoMoS$_2$ has been realized by experiment [5]. If we consider the poor Co environment, the $E_{Co}$ can be roughly estimated by the energy of a single Co atom. The formation energy defined as $\Delta E_3 = (E_{(MoS_2)_2Co} - E_{(MoS_2)_2} - E_{Co})/n$ is about -0.138 eV/atom. Therefore, the controlling of the concentration of Co may be very critical for the experimental synthesis of the multiferromagnetic Co(MoS$_2$)$_2$.

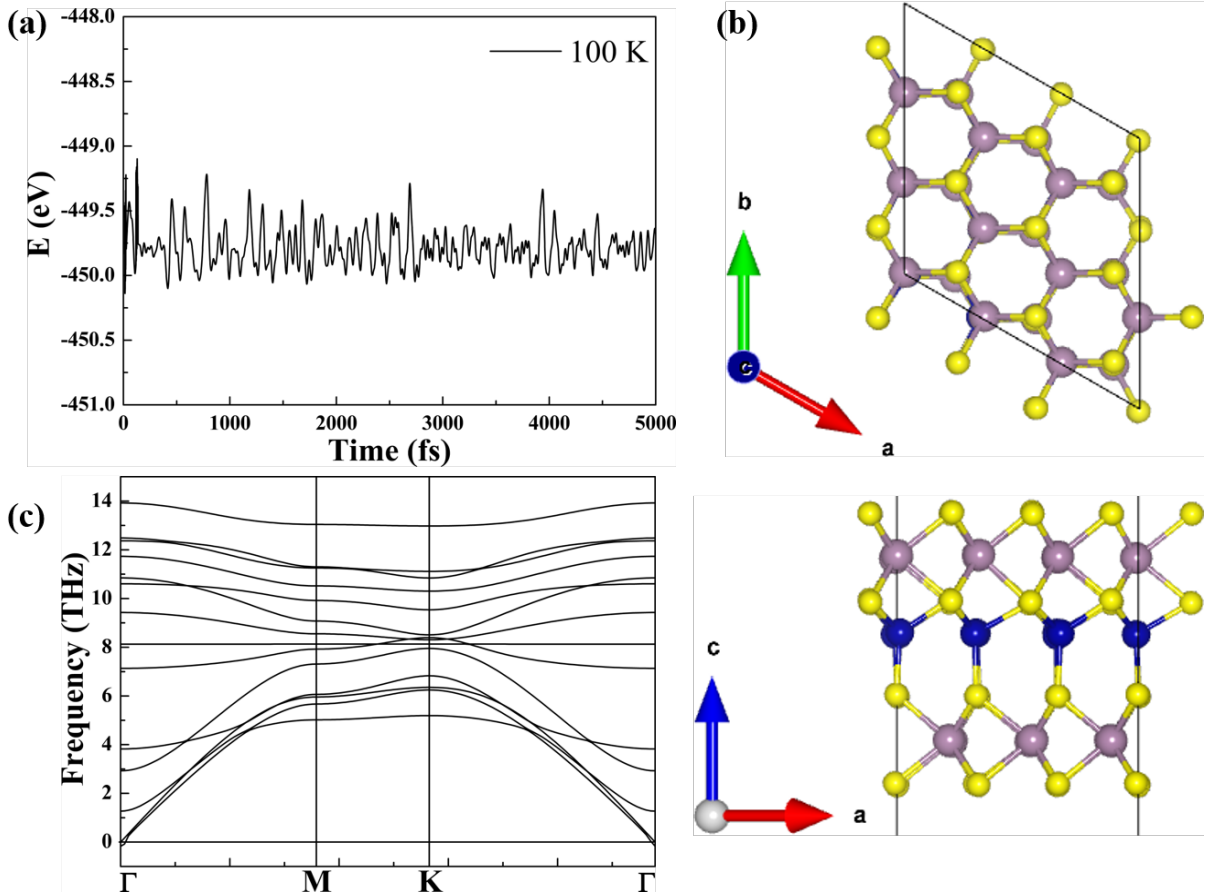

FIG. S2. (a) Evolution of the total energy with respect to molecular dynamics simulation step and (b) snapshot of the equilibrium structure for Co(MoS$_2$)$_2$ after simulation time of 5 ps. The *ab initio* molecular dynamic (AIMD) simulation is carried out with a time step of 1 fs by using a 3×3×1 supercell under a Nose-Hoover thermostat at 100 K. (c) The phonon band structure without imagine frequency is displayed.

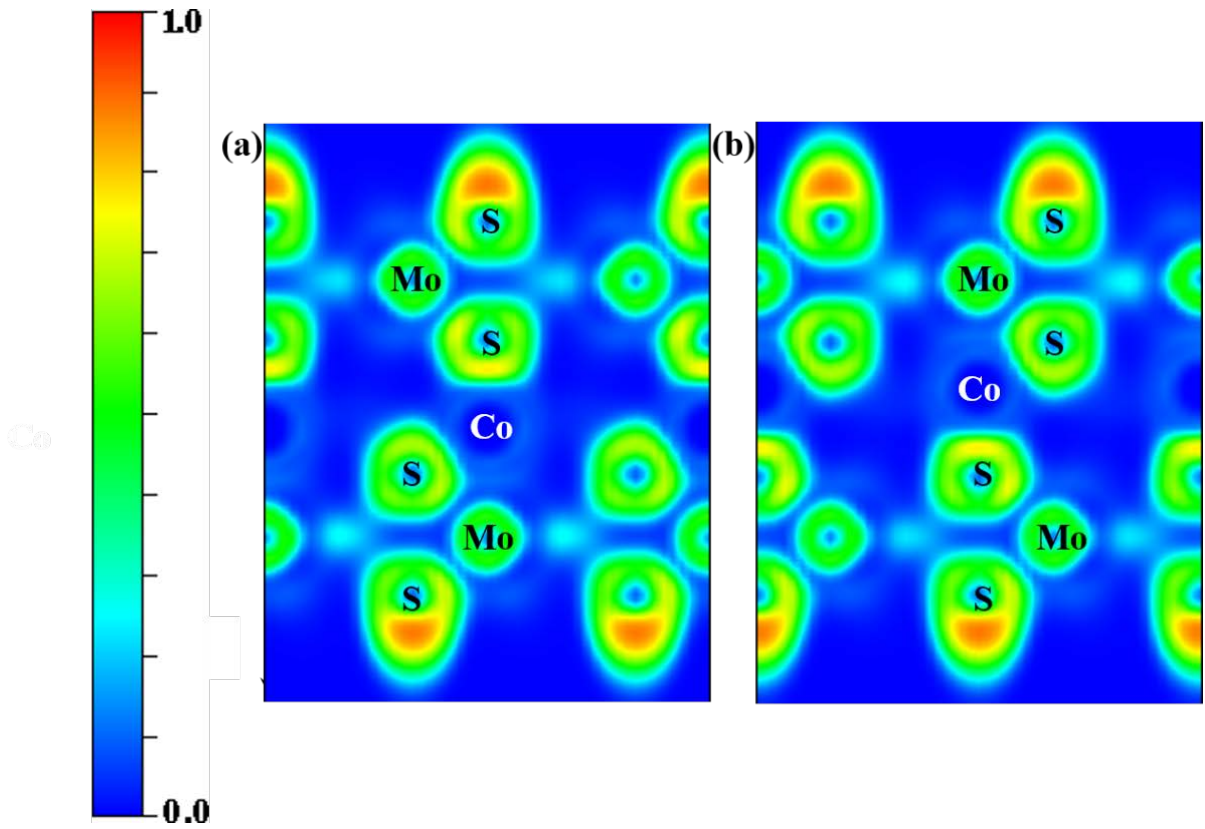

FIG. S3. The electron localization function of the $(2\bar{1}0)$ plane of (a) FE1 and (b) FE2 phases of Co(MoS$_2$)$_2$ monolayer.

## 2 Calculations of the magnetic interaction parameters *J*, *K,* and *D*.

The magnetic anisotropy energy (MAE) $K$ is obtained by calculating the energy difference between in-plane [100] and out-of-plane [001] magnetization orientations:

$$E_{MAE} = E_{100} - E_{001} \quad (S1)$$

To calculate the Heisenberg exchange interaction *J*, 2×1×1 supercell is built with the collinear ferromagnetic (FM) and antiferromagnetic (FM) spin configurations (Figure S3). According to Eq. (1), total energies of FM and AFM are:

$$E_{FM} = E_0 - 6J - 2K \quad (S2)$$

$$E_{AFM} = E_0 + 2J - 2K \quad (S3)$$

where $E_0$ is the total energy of the system without magnetic interactions. The DMI term is zero since spins in FM and AFM configurations are collinear. Therefore, parameters *J* can be determined using Eqs. (S2) and (S3).

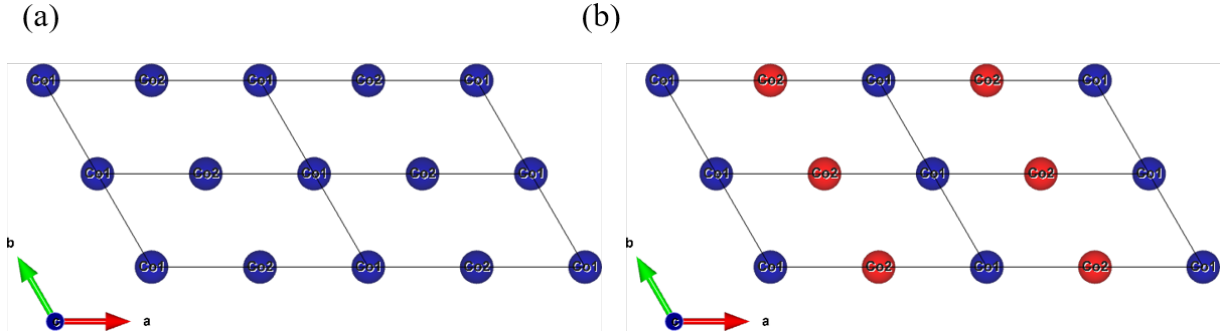

FIG. S4. Spin configuration of (a) FM and (b) AFM of Co atoms for the calculation of Heisenberg exchange interaction *J*. The blue and red represent spin up and down orientations, respectively.

The DMI energy between normalized spins between the nearest Co atoms can be written as:

$$E_{DMI} = \sum_{i,j} \boldsymbol{d}_{i,j} \left( \boldsymbol{S}_i \times \boldsymbol{S}_j \right) \quad (S4)$$

The different vertical distances between the intercalated Co atoms and two TMD layers breaks the structural inversion symmetry and thus gives rise to DMI interaction. The system under consideration possesses $C_{3v}$ symmetry with a three-fold rotation axis at the center of the three nearest Co atoms. Therefore, the mirror plane along the rotation axis perpendicularly passes through the bond between two nearest Co atoms. According to the Moriya symmetry rules, the DMI can be expressed as:

$$D_{i,j} = d_\parallel (\hat{u}_{i,j} \times \hat{z}) + d_{ij,z}\hat{z} \quad (S5)$$

where $\hat{u}_{i,j}$ is the unit vector between site $i$ and $j$, $\hat{z}$ indicates the unit vector normal to the plane. $d_\parallel$ is the in-plane component of $D_{i,j}$ that can be calculated using the chirality-dependent total energy difference approach. We calculated the energies difference between clockwise (CW) and anticlockwise (ACW) spin configurations and thus $d_\parallel = (E_{acw} - E_{cw})/12$. The out-of-plane component $d_{ij,z}$ can be neglected in the system due to sign variation of $d_{ij,z}$ for the six nearest Co atoms.

# 3 Magnetic parameters and spin textures of the field-induced transition structures

To investigate the evolution of the spin textures under an electric field, we calculated the magnetic parameters $K$, $J$, and $D$ of each field-induced transition structure between the FE1 and FE2 switch as functions of polarization (Figure S5).

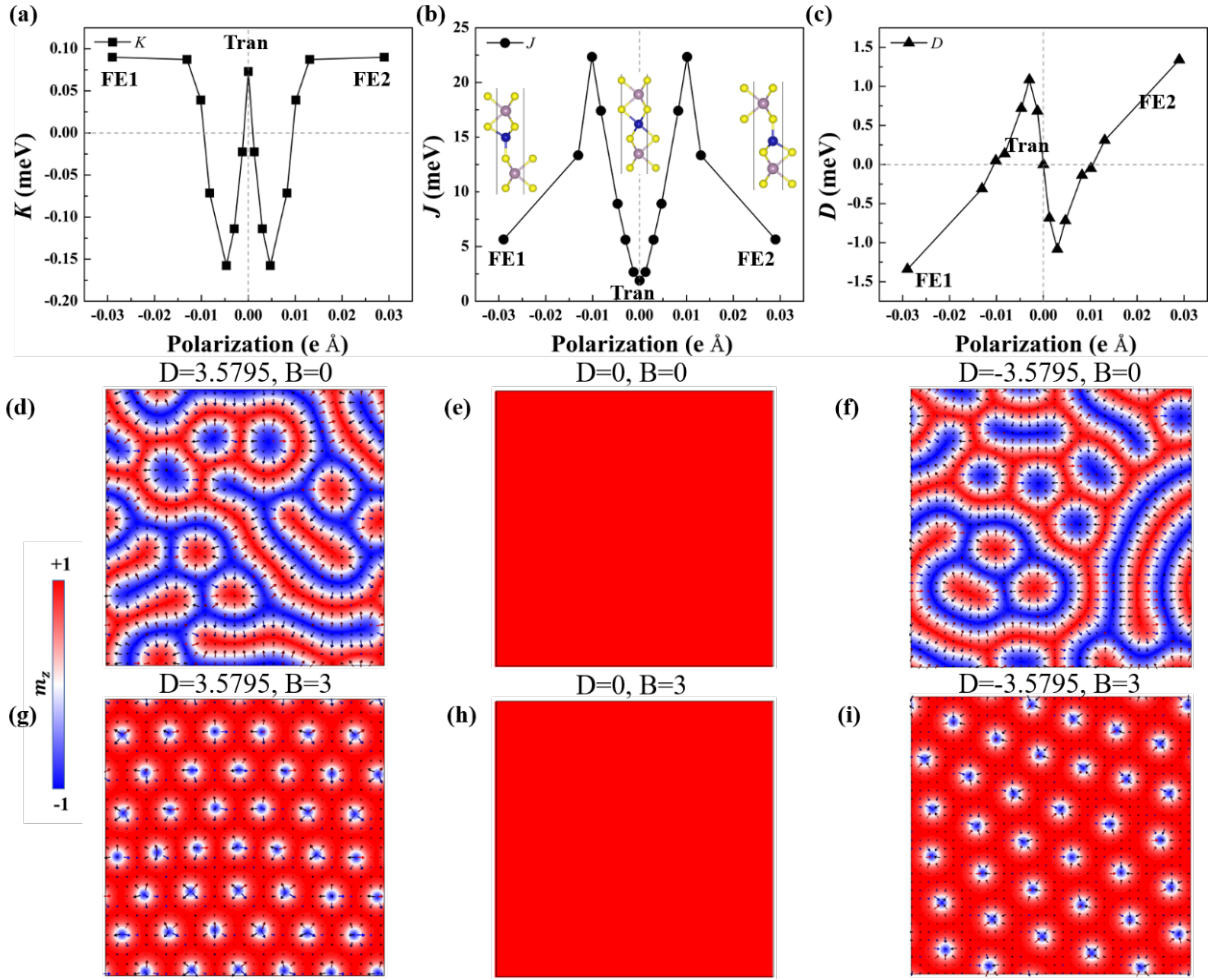

FIG. S5. The evolution of (a) $K$, (b) $J$, and (c) $D$ parameters as a function of polarization. The spin textures of the ferroelectric phase FE1, the paraelectric transition phase Tran, and the ferroelectric phase FE2 under 0 T (d-f) and 3 T (g-i) external magnetic field applied, respectively.

The magnetic anisotropy parameter $K$ [FIG. S5(a)] shows that the magnetic anisotropies of the states near the FE1, FE2, and the paraelectric transition (Tran) phase are out-of-plane, while the other transition states are in-plane. The positive Heisenberg exchange coupling parameter *J* [FIG. S5(b)] indicates that the ferromagnetic state is always maintained during the electric-field induced transition between FE1 and FE2. The DMI parameter *D* in Figure S5c changes the sign as the polarization is reversed. To investigate the effect of the external electric field on spin textures, we carried out the micromagnetic simulations of FE1, FE2, and Tran phases shown in FIG. S5(d-i). Under zero magnetic field, the FE1 phase with positive polarization has an anticlockwise DMI, the ferromagnetic warm-like domain walls are found. As the electric field increases, the polarization and the DMI value decrease to zero in the field-induced transition phase Tran with central inversion symmetry, the system prefers ferromagnetic states. Further increasing electric field will switch the polarization by forming the FE2 phase. The FE2 phase with negative polarization has a clockwise DMI resulting in ferromagnetic warm-like domain walls with opposite chirality compared with the FE1 phase. Under external magnetic field of 3 T, skyrmions are formed in FE1 and FE2 phases. Micromagnetic simulations indicate that the external electric field will reverse the chirality of the skyrmions. Therefore, the external electric field can realize the switch between two spin textures with opposite chirality through changing the sign of DMI by inducing the transition from FE1 to FE2 phase.

## 4 The impact of effective Coulomb interaction $U_{eff}$

The influence of $U_{eff}$ on the magnetic parameters ($K$, $J$, and $D$) which can define the configuration of spin textures are calculated and shown in Fig. S6.

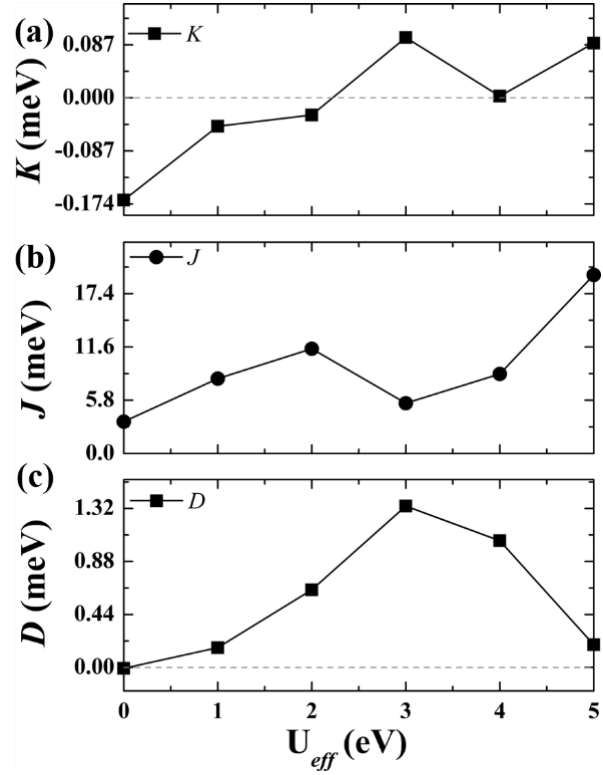

FIG. S6. The magnetic parameters (a) $K$, (b) $J$ and (c) $D$ under different $U_{eff}$.

It follows that the Heisenberg exchange coupling parameter J overall increases as a function of $U_{eff}$, while MAE changes from in-plane to out-of-plane and the DMI parameter first increases and then decreases. Here we choose $U_{eff}$ = 3 eV since the total energy of cobalt under oxidation environment can be accurately calculated [6,7].

## 5 Magnetic parameters of two ferroelectric phases in micromagnetic simulations

TABLE SI Calculated Heisenberg exchange coupling $J$, effective magnetic anisotropy energies $K_{eff}$, DMI parameters $D$, and saturation magnetization $M_s$ of the ferroelectric $Co(MoS_2)_2$ phases FE1 and FE2 with opposite polarization in micromagnetic simulations.

|  | $A$ (pJ/m) | $K_{eff}$ (MJ/m$^3$) | $D$ (mJ/m$^2$) | $M_s$ ($10^5$ A/m) |
|---|---|---|---|---|
| **FE1(P↑)** | 4.8611 | 0.4971 | 3.5795 | 4.09786 |
| **FE2(P↓)** | 4.8611 | 0.4971 | -3.5795 | 4.09786 |

# 6  Other potential TMDs based multiferroic materials

TABLE SII The optimized lattice constant, magnetic parameters ($J$, $d$, $K$ and $\mu_{Co}$) and electric polarization of Co(MoSe$_2$)$_2$, Co(MoTe$_2$)$_2$ and Co(WS$_2$)$_2$.

|  | a (Å) | $J$ (meV) | $d$ (meV) | $K$ (meV) | $\mu_{Co}$ ($\mu_B$) | P (e Å) |
|---|---|---|---|---|---|---|
| **Co(MoSe$_2$)$_2$** | 3.372 | 6.093 | 0.440 | 0.104 | 1.60 | 0.012 |
| **Co(MoTe$_2$)$_2$** | 3.629 | 7.036 | 1.044 | -0.239 | 1.50 | 0.012 |
| **Co(WS$_2$)$_2$** | 3.216 | 4.815 | 2.632 | 0.057 | 1.19 | 0.023 |

**Supporting information References**

[1] Y. Wang, J. Lv, L. Zhu, and Y. Ma, Comput. Phys. Commun. **183**, 2063 (2012).

[2] D. Zhou, D. V. Semenok, D. Duan, H. Xie, W. Chen, X. Huang, X. Li, B. Liu, A. R. Oganov, and T. Cui, Sci. Adv. **6**, eaax6849 (2020)

[3] M. Yu, X. Liu, and W. Guo, Phys. Chem. Chem. Phys. **20**, 6374 (2018).

[4] P. Jiang, L. Kang, X. Zheng, Z. Zeng, and S. Sanvito, Phys. Rev. B **102**, 195408 (2020).

[5] M. Huang, J. Xiang, C. Feng, H. Huang, P. Liu, Y. Wu, A. T. N'Diaye, G. Chen, J. Liang, H. Yang, J. Liang, X. Cui, J. Zhang, Y. Lu, K. Liu, D. Hou, L. Liu, and B. Xiang, Direct Evidence of Spin Transfer Torque on Two-Dimensional Cobalt-Doped MoS2 Ferromagnetic Material, ACS Appl. Electron. Mater. 2, 1497 (2020).

[6] L. Wang, T. Maxisch, and G. Ceder, Phys. Rev. B **73**, 195107 (2006).

[7] C. Le, S. Qin, and J. Hu, Sci. Bull. **62**, 563 (2017).